\documentclass[%
 reprint,
 amsmath,amssymb,
 aps,
pra,
]{revtex4-1}

\usepackage{graphicx}
\usepackage{dcolumn}
\usepackage{bm}
\usepackage{color,soul}

\begin{document}

\preprint{APS/123-QED}

\title{Quantifying Urban Road Network Vulnerability and Resilience to Attacks}
\author{Skanda Vivek}
\email{skanda.vivek@gmail.com}
\affiliation{School of Science and Technology, Georgia Gwinnett College, Lawrenceville, GA 30043}

\date{\today}

\begin{abstract}
The rise of connected and autonomous vehicles, combined with the proliferation of IoT and connected surfaces, lead to the emergence of novel complex cyber risks. Lack of encryption and authentication in internal vehicular networks are widely recognized as cause for concern by cybersecurity experts, automobile, and OEM manufacturers. This concern has only been growing with the increase in cybersecurity incidents and demonstrations showing different vehicular vulnerabilities, making it nearly impossible to completely secure vehicles against cyber-attacks. Of particular concern is the potential for large-scale vehicular cyber-attacks to cascade to transportation networks, which are the lifeline of cities.  Here, we develop a framework based on complex network theory, traffic flow, and new data based technologies to quantify the vulnerability of city-scale transportation to cyber-attacks. Application of our framework to the road network of Boston reveals that targeted attacks on a small fraction of nodes leads to disproportionately larger disruptions of routes. We develop an early-detection framework to quantify real-time risk based on gathering multidimensional traffic flow, incident, and social media data sets. Our results illustrate an effects based approach to transportation cyber-defense, through informed, intelligent vehicular agents. 

\end{abstract}
\maketitle


\section{Introduction}
\begin{figure*}[hbt!]
\centering
\includegraphics[width=\textwidth]{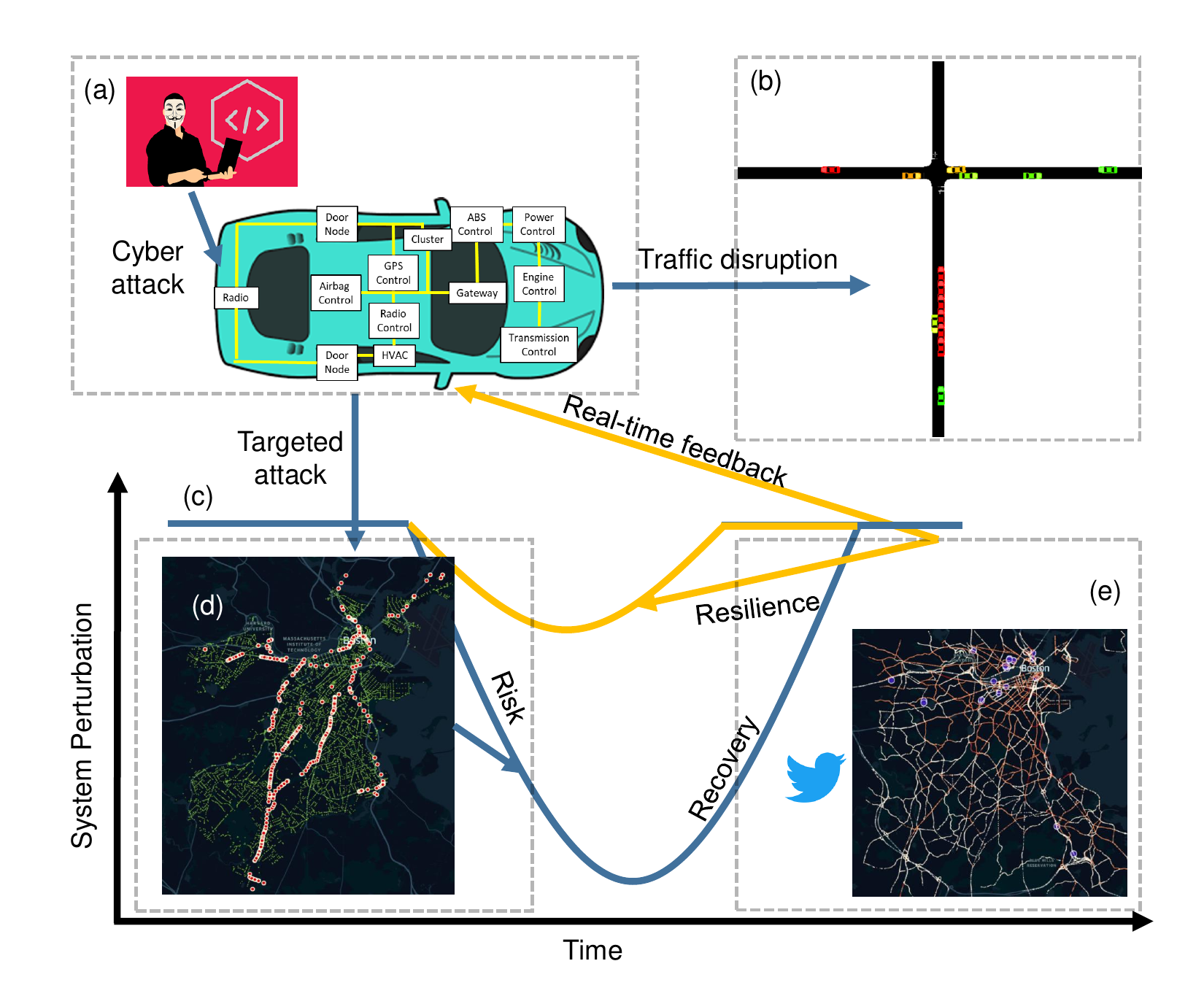}
\caption{Tightly coupled vehicle cyber and city transportation networks. (a) Cyber-attack on intra vehicular networks impacting vehicle functioning. (b) Additional impacts on vehicular traffic. (c) Illustration of system perturbation following a large-scale hack and subsequent recovery (blue). Potential for improved resilience through reduced perturbation and faster recovery following early detection through novel, multiple data sources (yellow). (d) Cascading impacts on city-scale transportation networks (e) Potential for building in resilience through early-warning from multiple traffic flow, incident, and social media data sources. }
\label{fig:fig1}
\end{figure*}

Two recent trends of (i) increasing connectivity and (ii) tight coupling arising from cyber-physical complexity drive the potential for cyber-attacks to result in large scale disruptions or in some cases; complete failure of critical infrastructures. It is estimated that 20\% of vehicles in the US can connect to the internet, and this market share is growing every year. Along with the growth in connected vehicle surface area, comes the risk of vehicular hacking. There has been a growth in incidents where hackers have demonstrated that they can gain control of safety-critical vehicle functions such as brake, engine, and steering~\cite{koscher2010,checkoway2011,miller2015}. Various surfaces have been found for example through vehicle entertainment control unit~\cite{miller2015}, spoofing vehicle sensors like LIDAR~\cite{cao2019}, and poorly authenticated 3rd party apps that are linked to critical vehicle safety functions~\cite{eiza2017}. Ultimately, it becomes very hard to ensure that vehicles are completely secure against vulnerabilities. 

Studies in complex networks  have resulted in approaches to evaluate the fragility of connected networks to disruptions and potential for cascading disruptions~\cite{albert2000,motter2002,motter2004,zhao2016}. In recent years, network based approaches have been applied in the context of existing critical infrastructure disasters~\cite{duan2019}, including the power grid~\cite{buldyrev2010}, road networks~\cite{ganin2017,zhu2019}, and aircraft delays~\cite{fleurquin2013}. However, the coupling between cyber and the physical realm of transportation leads to the potential for complex risks, that are so far largely unknown. 

Arguably the greatest potential for disruption from a large-scale cyber attack on connected vehicles lies in the physical realm of transportation. One of the greatest security threats for connected vehicles are fleet wide hacks, stated by Tesla CEO Elon Musk in 2017~\cite{Musk2017}. While it is hard in the absence of data, to identify how a cyber attack might effect vehicle functions, several studies and hacking demonstrations have shown that this could lead to collisions~\cite{amoozadeh2015} or cause vehicles to enter a safe mode~\cite{vivek2019,parkinson2017,Cal2018-article,Cal2018-rules}, decreasing the likelihood for collisions but increasing local local traffic jams around compromised vehicles. Other studies have shown that hackers could intentionally shutdown vehicles~\cite{franceschi2019}.

Here, we develop a framework to quantify and early-detect the impacts of vehicular cyber attacks on transportation networks. Our work leverages on complex network theory, and concepts from tightly coupled complex systems~\cite{perrow2011}. First, we develop a theory to quantify the vulnerability of transportation to targeted cyber attacks (Fig.~\ref{fig:fig1}a). Through simulations on representative grids, we observe ensuing cascading traffic jams around blocked vehicles, shown in red in Fig.~\ref{fig:fig1}b. These observations help inform a theoretical framework that we apply to quantify city scale disruptions through cell phone location data sets in combination with spatial road networks. Application of our methods on the street network of Boston identifies the most vulnerable nodes through a measure of network centrality (Fig.~\ref{fig:fig1}d). Finally, we develop a novel framework for early-detection of cyber attacks through multidimensional incident, flow, and social data sets, that serve as a pulse of modern societies (Fig.~\ref{fig:fig1}e). 

Figure~\ref{fig:fig1}c demonstrates the system perturbation in the absence (blue) of detailed information, and the presence (yellow) of real-time information. Through our framework of near real-time situational awareness quantified from multidimensional data 
sources, various stakeholders such as automobile manufacturers, emergency managers, drivers, and vehicles themselves can make informed decisions that mitigate propagation of perturbation, and speed up recovery. In the broader context, our work conceptualizes an effects based approach to identify critical infrastructure assets most vulnerable, and build resilient infrastructure networks to cyber attacks.

\section{Results}
\subsection{Quantifying post-hack traffic disruptions on model grids}
To quantify post-hack traffic disruptions, we do traffic simulations using the Simulation of Urban Mobility (SUMO) platform~\cite{sumo} on representative grids. In the simulations, vehicles enter the grid at a fixed rate, and exit once they have finished their route. Routes are chosen randomly, and the rate of entry corresponds to uniform density of 6 vehicles/km/lane in free flow conditions. Each edge is bidirectional, and each road has 1 lane. Roads were blocked at the center, and in the event of road blockage, vehicles could not pass through the road. An initially blocked road leads to a traffic jam. Eventually, this traffic jam cascades to adjacent roads, blocking vehicles using those roads in turn, and this cascade continues until the whole grid is blocked by vehicles. 

\begin{figure}[hbt!]
\begin{centering}
\includegraphics[width=\columnwidth]{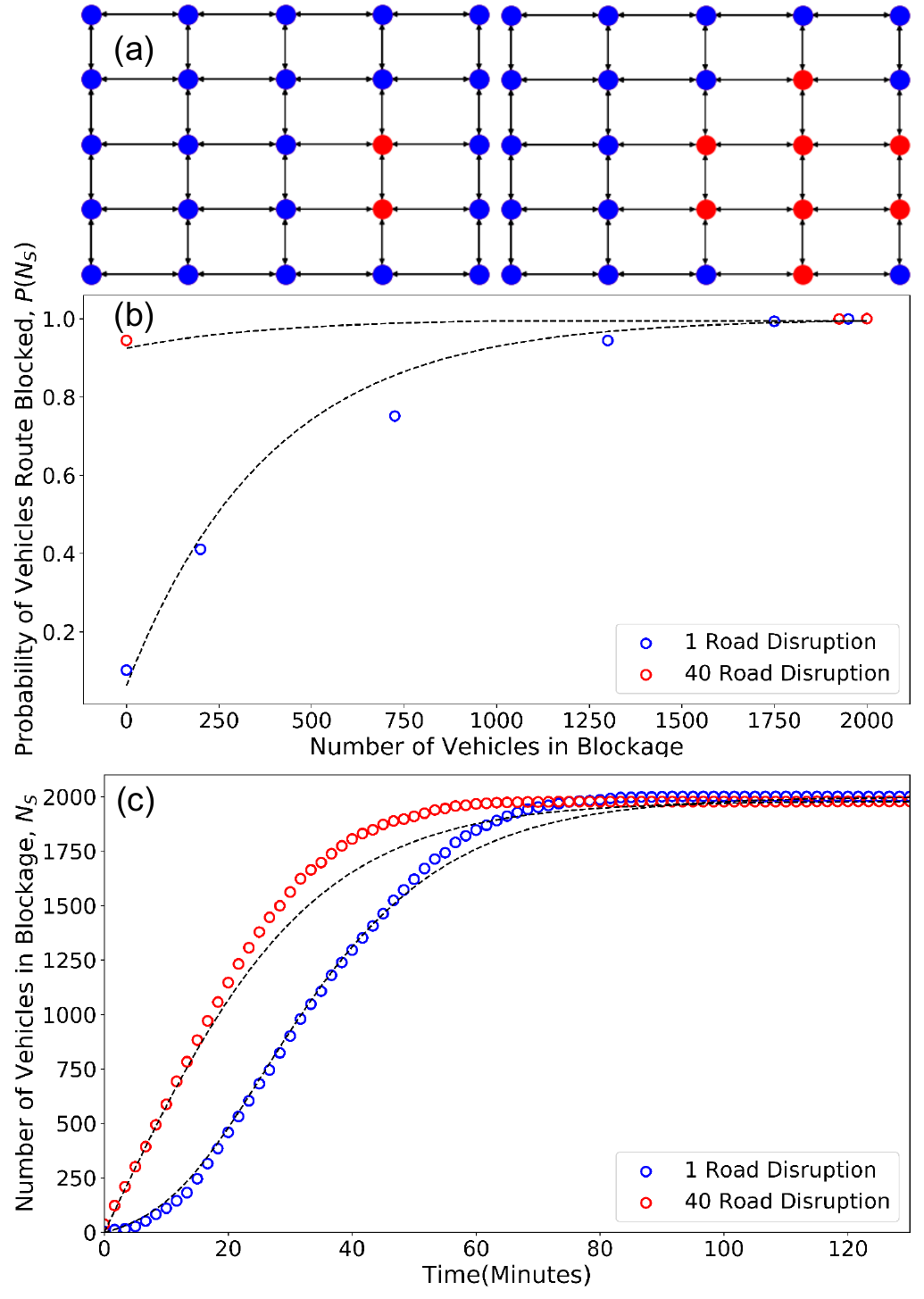}
\caption{Simulations and analytical calculations of post-hack vehicular traffic on grids. (a) 5x5 grid, with intersections denoted by nodes and roads by edges. Cascade of traffic jams from initial road (adjacent intersections in red, left) to neighboring roads (right). (b) Probability kernel $P(N_S$) showing the probability that vehicles routes are blocked with number of stopped vehicles. Initially, $\sim$ 6\% of routes are blocked due to a single road disrupted and $\sim$ 90\% of routes are blocked
when half the roads are disrupted. Dashed lines are theoretical prediction. (c) Growth of stopped vehicle traffic jam with time, colors correspond to different numbers of initially blocked roads along with theoretical formula (black dashed lines). }
\label{fig:fig2}
\end{centering}
\end{figure}

To understand how an initially blocked road leads to cascading network failure through traffic jams of stopped vehicles, we use NetworkX~\cite{networkx}, a python based network analysis package. NetworkX treats roads as edges and intersections as nodes. Figure~\ref{fig:fig2}a shows an example scenario in a 5x5 network, where an edge is initially blocked. The 2 nodes connecting the edge are marked in red (Fig.~\ref{fig:fig2}a, left). Each road is 200m long and can contain a maximum of 25 vehicles end to end. Figure~\ref{fig:fig2}a (right) shows the cascade when traffic jams spill over from the initially blocked edge to neighboring edges, whose corresponding nodes are marked in red. 

Using the route information in the corresponding SUMO simulation, we quantify the fraction of routes inaccessible if roads are blocked. From this we obtain the probability of a vehicle route being blocked as a function of number of vehicles in the blockage, $P(N_S)$ (Fig.~\ref{fig:fig2}b). Note that in the case of 1 edge disruption, $\sim$6\% of vehicle routes are impacted before there are any vehicles in the blockage. Whereas in the case of 40 edges initially disrupted, $\sim$90\% of vehicle routes are impacted before there are any vehicles in the blockage. This initial disruption cascades to the entire grid as more vehicles enter the blockage. The dashed black lines denote the theoretical formula from probabilistic arguments, with the assumptions are that all edges are equally probable for routes, and independent of each other. The theoretical formula for $P(N_S)$ is:

\begin{equation}
    P(N_S) = 1-\left (1-\frac{N_R}{N_{T}} \right)^{\textrm{min}(h_0+N_S/25,N_T)}
\end{equation}

Here, $N_R$ denotes the average number of edges in a route (which is 5 for 5x5 networks), $N_T$ is the total number of edges in the network (80 bidirectional roads in a 5x5 network), $h_0$ is the initial number of disrupted roads, and $N_S$ is the number of vehicles in the blockage. Since there are a maximum of 25 vehicles possible in each 200m edge (8m separation between vehicles), $N_S/25$ represents the number of additional roads blocked from cascading vehicle blockages. However, this is limited by the maximum number of edges that can be blocked ($N_T$), given the finite size of grids. Circles are measurements of inaccessible routes from SUMO simulations, which are well represented by the theoretical formula.

Using this probability, we develop a theoretical formula to predict how the number of stopped vehicles grows in time, as a function of $P(N_S)$, rate at which vehicles enter the simulation $r$, and the maximum capacity of vehicles in the grid, K:

\begin{equation}
    \frac{dN_S}{dt} = r\cdot P(N_S)\cdot \left (1-\left(\frac{N_S}{K}\right)^2\right) 
\end{equation}

In equilibrium, $r$ is related to the time spent by vehicles and number of vehicles in the simulation at any given moment as $r*T=N$. Further, $T=N_R*200/v$, where $v$ is the average velocity of vehicles in the simulation in m/s and each road is 200m long. $r\cdot P(N_S)$ denotes the contribution of vehicles entering the simulation to the rate of growth of vehicle blockage. In the limit no routes are blocked, $P(N_S)$ is 0, and there is no growing blockage. When all routes are blocked, $P(N_S)=1$ and all vehicle routes are blocked, thus all incoming vehicles enter the blockage. 

The $(1-N_S/K)$ contribution is due to the finite vehicle capacity of the grid. As more of the grid is occupied by traffic jams, vehicles cannot enter the simulation at their intended origin, which diminishes the rate of growth of stopped vehicle blockage close to grid carrying capacity. The $(1+N_S/K)$ term is because existing vehicles in the simulation cannot reach their intended destination due to the growing traffic jam, and instead attach to the growing vehicle blockage quicker than their intended destination. Thus, the $(1-N_S/K)^2$ term accounts for vehicles not able to leave depart from their origin due to the growing vehicle blockage, as well existing vehicles that attach to the blockage on their routes. Figure~\ref{fig:fig2}c shows that the theory matches traffic simulation results quite well for different initial numbers of roads blocked. We also confirmed this for simulations of different grid sizes. 

\subsection{Applying our theoretical framework of transportation disruption to the Boston road network}

In our SUMO simulations on grids, routes were chosen randomly, and each road has the same importance as every other route. However, in city road transportation networks certain routes like central highway systems are more important than others. To account for this, we use information of city vehicle trip routes, explained in the following section.
For Boston route information, we use data from a previous study which obtained the origin (O) and destinations (D) for individual travelers during peak 7:30-8:30 AM rush hour~\cite{colak2016}. The OD information was obtained though tracking of cell phone records of a target population during that period. 

To build the urban road network of Boston, and analyze the relative importance of roads in the context of traveller routes, we used OSMnx~\cite{osmnx}, a python package that represents Open Street Maps information on streets as networks. The resultant map of Boston contained information about nodes and edges. Nodes correspond to intersections and edges to roads that connect intersections. While we do not have granular information about the routes taken, Open Street Maps provides limited information on the speed limits of roads. Using this information, we infer routes from OD data with the shortest free-flow travel times, based on Dijkstra's shortest path algorithm~\cite{dijkstra1959}.

\begin{figure}[hbt!]
\centering
\includegraphics[width=\columnwidth]{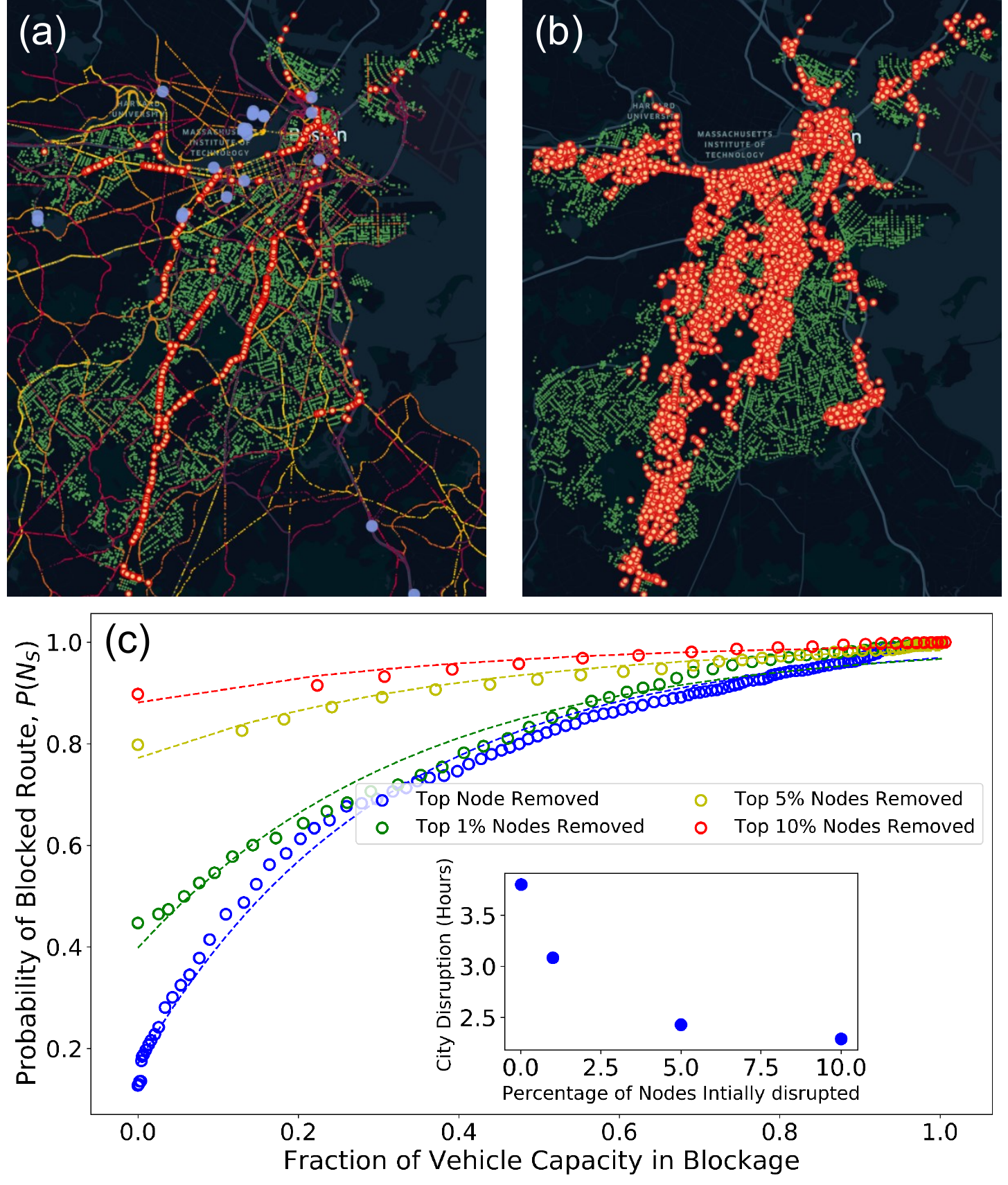}
\caption{Multilayered complex transportation and data networks, vulnerability to targeted cyber-attacks, and potential for early-detection through feedback. (a) Boston transportation networks with nodes from OSM (green), top 5\% important nodes (red circles), real-time traffic speeds data (colored lines), and real-time incidents (blue circles). (b) Cascading impacts of initially targeted 5\% important nodes to the rest of the city. (c) Probability of shortest time routes being blocked vs fraction of vehicle capacity in the blockage. Colors correspond to different initial fractions of roads disrupted and dashed lines are inverse exponential fits. Blue circles in the inset are time for half of Boston to be filled with stopped vehicles vs initial scale of disruption.  }
\label{fig:fig3}
\end{figure}

In order to understand the fragility of urban routes to targeted disruptions, we apply our framework, in particular the probability kernel $P(N_S)$, to real trips in Boston. For Boston route information, we use data from a previous study which obtained OD pairs from travelers during peak 7:30-8:30 AM rush hour~\cite{colak2016}. The OD information was obtained though tracking of cell phone records of a target population during that period. To analyze Boston road networks, we used OSMnx~\cite{osmnx}, a python package for working with Open Street Maps. The resultant map of Boston contained information about nodes and edges. Nodes correspond to intersections and edges to roads that connect intersections. While we do not have granular information about the routes taken, we obtained the shortest time paths corresponding to the OD data in the Boston road networks. For each route, we obtained all the nodes in sequence, corresponding to the shortest time path. 

In regular taffic conditions, one would expect the various map applications to show the shortest time paths. Figure~\ref{fig:fig3}a shows the top 5\% of nodes (red circles) in the Boston road network. Visually, it appears most of the nodes are clustered around highways like I-90 and central streets such as Washington street. When these roads are initially disrupted, this cascades to adjacent roads as can be seen in Figs.~\ref{fig:fig3}b.

The importance of a node is measured through the fraction of shortest time routes passing through the node, out of all routes as below:

\begin{equation}
    g(v)=\frac{\sum_O\sum_D \delta_{OD,v}}{\sum_O\sum_D \delta_{OD}}
    \label{eqn:centrality}
\end{equation}

where $g(v)$ is the importance of the node $v$, $\delta_{OD}=1$ when a route has origin $O$ and destination $D$, and $\delta_{OD,v}=1$ when a route with origin $O$ and destination $D$, passes through node $v$. The summations are done over all origin-destination pairs. Depending on the scenario, this formula can be modified to reflect other measures of centrality such as betweenness, which have been applied to road networks~\cite{freeman1977,batista2015}.

Surprisingly, we find that most of the shortest time routes from the OD pair data pass through a very small fraction of nodes. Figure~\ref{fig:fig3}c shows that 10\% of all shortest time routes passes through the single most important node, and almost 80\% of routes pass through the top 5\% nodes in Boston. This shows that the Boston routing network is extremely fragile to a targeted disruption of a few essential nodes. Once vehicles start piling up, the situation cascades to city scale failure. In the case of the single most important node being shutdown, a traffic jam involving 10\% of the Boston vehicle capacity would result in 30\% of shortest time routes being blocked. To extract $P(N_S)$, we fit an inverse exponential to our $P(N_S)$ measured from shortest time paths, where $p_0$ and $a$ are fitted, $p_0$ corresponds to the initial disruption at $N_S=0$ i.e. fraction of shortest paths initially inaccessible.

\begin{equation}
    P(N_S)  = (1-p_0)(1-\exp(-a N_S))+p_0\\
    \label{eqn:bos1}
\end{equation}

We apply our extracted kernel $P(N_S)$ to reveal the dynamic cascade. For this, we obtain the maximum number of vehicles possible in Boston ($K_{Bos}$) by dividing the total road length* number of lanes in m/8; assuming a uniform spacing of 8 m between vehicles at maximum capacity which corresponds to a maximum density $\rho_{max}=125$ vehicles/km/lane. To find the rate of vehicles entering we use the information of ODs from rush hour obtained from cell phone data. We apply these values in the formula below:

\begin{equation}
    \frac{dN_S}{dt}  = r_{rush}\cdot P(N_S)\cdot \left (1-\left(\frac{N_S}{K_{Bos}}\right)^2\right) 
    \label{eqn:bos2}
\end{equation}
 $r_{rush}$ is the number of trips per second, calculated from the number of trips during rush hour. Figure~\ref{fig:fig3}c (inset) shows that time at which half of Boston is filled is between 2-4 hours, depending on the percentage of nodes initially disrupted. Interestingly, above 5\% of nodes hacked, there is a point of \textit{diminishing return} above which more nodes disrupted does not speed up the ensuing cascade.

\subsection{Early-detection framework for cyber attacks through multidimensional data}
Figure~\ref{fig:fig3}a shows multiple datasets on traffic speed and incidents obtained from the HERE traffic API~\cite{here}, overlayed on the spatial map of Boston, along with the top 5\% of nodes  shortest routes, as obtained from the centrality metric defined in Eqn.~\ref{eqn:centrality}. Real-time traffic information by HERE or other providers, could be used to measure city-scale perturbations of traffic from regular conditions, such as due to cyber attacks. In addition, its been shown that behavior on social media e.g. Twitter also contain useful signatures of traffic anomalies~\cite{pan2013,giridhar2016}. The equations below represent a novel framework to combine disparate data sets for early detection of cyber attacks.

\begin{equation}
    R[\vec{X_j}]=\sum_{Dim}\sum_{i}R[\vec{x_i}]\delta_{G(\vec{x_i}),X_j}
    \label{eqn:bos3}
\end{equation}

\begin{equation}
    C\cdot R=\sum_j R[\vec{X_j}]e^{- \lVert{\vec{X_j}-B[\vec{X_j}]}\rVert}
    \label{eqn:bos4}
\end{equation}
Here, $R[\vec{x_i}]$ denotes the risk due to incident $i$ located at location $\vec{x_i}$. $R[\vec{x_i}]$ can be quantified based on the severity of the incident. For example, HERE traffic incident data provides information on the type of incident as well as criticality level of the incident that can be useful in evaluating the risk level of the incident. $G(\vec{x_i})$ is the mapping of that incident to the closest intersection(node) location, $X_j$  on the Boston road network. The sum over $i$ denotes all the incidents that can be mapped to the node $j$, and the sum over $Dim$ denotes the sum over all the different data sets such as flow, incident, twitter, etc. $R[\vec{X_j}]$ gives the total risk at node $j$.

In order to compute city scale risk, it is important to identify the subset of nodes that are critical and consequently at risk for targeted disruption. These nodes are represented as basis set $B$, for example the top 5\% of nodes in the Boston road network. Risk at node $j$ is then weighted according to the absolute distance from the closest corresponding basis node $B[\vec{X_j}]$ as $e^{- \lVert{\vec{X_j}-B[\vec{X_j}]}\rVert}$. A node that is at or very close to a critical node is weighted more than a node that is further away. Finally, the city scale risk, $C\cdot R$ is the sum of node risks, weighted by their respective distance from critical nodes. 
\begin{figure*}[hbt!]
\centering
\includegraphics[width=\textwidth]{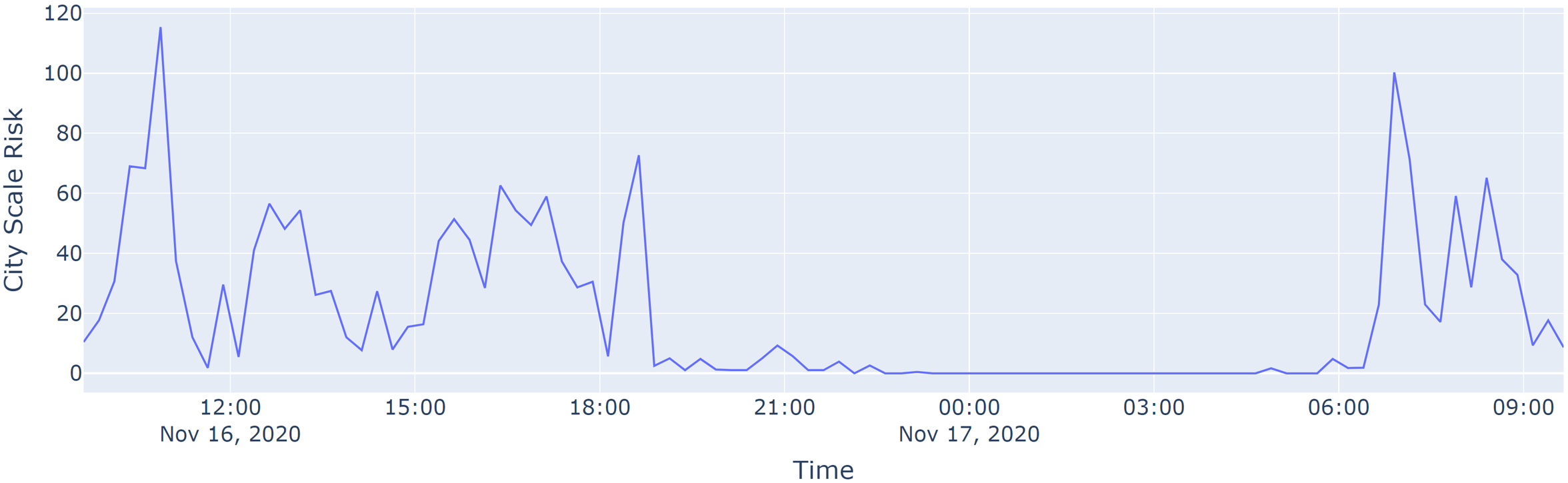}
\caption{City scale road transportation risk measured using HERE traffic flow data }
\label{fig:fig4}
\end{figure*}

As a proof of concept, we applied our framework on HERE traffic flow data collected from Boston during a 24 hour period in November 2020. As a proxy for locations at risk, we looked at locations on the urban network of Boston where traffic speeds were less than 0.25 times the speed limit, for example corresponding to a road speed limit 60 mph, if the road segment had a speed lower than 15mph it is considered an at risk location. Sudden drops in traffic speeds at multiple locations could indicate emergency disasters such as weather related extremes or future cyber incidents. To compute city scale risk, we chose the basis set of nodes $B$ as the top 5\% nodes as defined previously, on the Boston road network. We found the closest Basis node and corresponding distance from risk locations using the Haversine formula, that determines the great-circle distance between two points on a sphere given their longitudes and latitudes. For simplicity, all risk values $R[\vec{x_i}]$ were given a value of 1. By summing over exponentially weighted distances (in miles) of risk locations from the closest basis node corresponding to Eqn.~\ref{eqn:bos4}, we obtained the city scale risk.   Figure~\ref{fig:fig4} shows peaks in city  occur roughly corresponding at rush hour, and dips during night times. This gives a baseline of risk during nominal conditions in Boston, and significantly higher risks could be useful in detecting large-scale anomalies, with the greatest potential harm to urban road transportation.

\section{Discussion}

The United States Cybersecurity and Infrastructure Security Agency has identified 16 critical infrastructure sectors, so vital that their incapacitation would cause a debilitating effect on security, economic security, public health, or any combination of the above. Importantly, the Department of Homeland Security notes~\cite{kenneally2018} that there is insufficient knowledge of the risks facing critical infrastructure networks, as well as general reluctance of organizations to include a more complete evaluation of cybersecurity risks including external dependencies and downstream effects. In spite of the critical importance for considering multi-domain dependencies during cyber attacks, existing approaches are focused on narrowly defined asset levels such as software quality, malware analysis, and intrusion detection. Only a few studies account for risk arising from cross-dependencies between correlated networks during cyber attacks~\cite{linkov2013,vivek2019,harry2018}.

We have introduced a 3 layered framework for monitoring complex cyber risks through informed, intelligent agents which are: (i) Developing theoretical approaches to quantify multi-layered effects of cyberattacks. (ii) Applying these approaches to large-scale infrastructures through network maps, flow datasets, and centrality metrics.  (iii) Detecting system perturbations through network based risk metrics using multiple available data sources in near real-time. While we have illustrated these concepts in large-scale vehicular cyber-attacks and effects on the impending transportation network, our framework can be applied in general to other critical infrastructures. 

\section{Acknowledgements}

The author thanks David Yanni for discussions and Hannah Conner for supplementary code.

\bibliographystyle{unsrt}
\bibliography{skanda}

\end{document}